\documentclass[
  aps,
  pre,
  reprint,
  superscriptaddress,
  amsmath,
  amssymb,
  floatfix,
  longbibliography,
  showkeys
]{revtex4-2}

\usepackage[T1]{fontenc}
\usepackage[utf8]{inputenc}
\usepackage{lmodern}
\usepackage{graphicx}
\usepackage{microtype}
\usepackage[hidelinks]{hyperref}
\hypersetup{
  pdftitle={Edge-Reversal Noise and Percolation Transitions in Directed Networks},
  pdfsubject={Working research manuscript},
  pdfkeywords={directed networks, edge reversal, random orientation, percolation, strong connectivity}
}
\usepackage{xcolor}
\usepackage{comment}

\newcommand{\av}[1]{\left\langle #1\right\rangle}
\newcommand{\Pgf}{\widehat P}
\newcommand{\GSCC}{\mathrm{GSCC}}
\newcommand{\GIN}{\mathrm{GIN}}
\newcommand{\GOUT}{\mathrm{GOUT}}

\renewcommand{\doi}[1]{\href{https://doi.org/#1}{doi:\,\nolinkurl{#1}}}

\begin{document}

\title{Percolation under Edge-Reversal: \\Creation and Destruction of Strong Connectivity in Directed Networks 
}
\author{Mari\'an Bogu\~n\'a}
\email{marian.boguna\@ub.edu}
\affiliation{Departament de F\'isica de la Mat\`eria Condensada, Universitat de Barcelona, Mart\'i  i Franqu\`es 1, 08028, Barcelona, Spain}
\affiliation{Universitat de Barcelona Institute of Complex Systems (UBICS), Barcelona, Spain} 

\author{M. \'Angeles Serrano}
\email{marian.serrano\@ub.edu}
\affiliation{Departament de F\'isica de la Mat\`eria Condensada, Universitat de Barcelona, Mart\'i  i Franqu\`es 1, 08028, Barcelona, Spain}
\affiliation{Universitat de Barcelona Institute of Complex Systems (UBICS), Barcelona, Spain}
\affiliation{Instituci\'o Catalana de Recerca i Estudis Avan\c{c}ats (ICREA), Passeig Llu\'is Companys 23, E-08010 Barcelona, Spain}
\date{\today}

\begin{abstract}
We study how the connectivity of a directed network changes when each edge is independently reversed with probability $p$, while the underlying connections remain fixed. Reversal changes the in- and out-degrees, and we derive the exact transformation of the joint in--out-degree distribution and characterize its evolution under repeated reversal. However, the transformation does not rewire the network, and every edge still connects the same pair of vertices. As a result, the transformed degree distribution alone does not determine whether a giant strongly connected component exists, and a reconstruction based only on the transformed degrees can even place the network on the wrong percolation state. For the directed configuration model, we account for the preserved structure by separately tracking edges that retain or reverse their original direction, leading to a two-type branching process. From this formalism, we obtain explicit analytical criteria for the creation or destruction of a giant strongly connected component and illustrate these contrasting outcomes with empirical examples. Consistent with the predicted creation criterion, randomizing edge directions in a \emph{S. cerevisiae} transcriptional regulation network dramatically increases the size of the largest strongly connected component. In the San Diego directed road network, the degree moments satisfy the destruction criterion, and the largest strongly connected component decreases to virtually zero, although its decline occurs well below the predicted configuration-model threshold. These results provide a framework for assessing the sensitivity of mutual reachability to changes or uncertainty in edge directions and show that directional disorder can either enable or suppress global connectivity without changing the underlying connections.
\end{abstract}

\maketitle

\section{Introduction}
Directions encode function in many networks. They distinguish upstream from downstream regulation, supply from demand, sender from receiver, predator from prey, and one-way from two-way accessibility. A convenient way to probe how much a network depends on this information is therefore to perturb directions while keeping the underlying set of connections fixed. Here we study the simplest such perturbation: each directed edge is independently reversed with probability $p$ on the otherwise quenched network structure, meaning that every edge connects the same two vertices as before and only its direction may change.

The operation itself has clear precedents. Random orientations of a fixed graph have been studied in probabilistic graph theory~\citep{McDiarmid1981,Linusson2009,Narayanan2018}. Direction randomization has also been used as a null model in network science~\citep{Ahnert2008,Dominguez2014}, and partial edge reversals have been used to probe loop structure and global directionality in empirical networks~\citep{Itzhack2013,Rosen2014}. Independent reversal has been studied directly for the strongly connected component of divisor graphs~\citep{KimPhillips2026}. Related work has developed percolation
theory for random directed networks containing both unidirectional and
bidirectional edges~\citep{Boguna2005}, while strong-connectivity
transitions induced by one-way links have been studied in two-dimensional
lattice models~\citep{Verbavatz2021}. Our goal here is to develop an explicit theory for the effect of edge-reversal noise on degree statistics and, for directed configuration models, on percolation when the original edge pairings are preserved.

Generating functions provide a convenient way to describe how edge reversal changes the joint in--out-degree distribution, building on the standard treatment of arbitrary-degree directed random graphs~\citep{Newman2001}. Predicting percolation, however, requires more information in our setting because reversal preserves the original edge pairings. Reversing edge directions is therefore not equivalent to rebuilding a random graph from the transformed degree distribution. A current directed edge can arise in two different ways---as an original edge that retained its direction or as an original edge that was reversed. Following these two kinds of edges samples different excess-degree distributions and therefore gives different expected numbers of outgoing edges available for further exploration. We accordingly keep the two cases distinct in a two-type branching process. This construction builds on the standard branching-process description of giant components in sparse directed degree-sequence and configuration models~\citep{CooperFrieze2004,DonderwinkelXie2024}, and is closely related to multitype and correlated random-graph formalisms used when edge or stub classes carry information not contained in the degree distribution alone~\citep{Soderberg2003,Boguna2005,Allard2015}.

For a directed configuration model, keeping track of the two possible arrival types---through a retained or a reversed edge---leads to a two-type branching description of percolation. Its largest eigenvalue determines the percolation threshold for a giant strongly connected component to exists. Because reversal changes the relative contribution of the two arrival types to the largest eigenvalue, the resulting threshold can move either upward or downward as reversal noise increases, so the same perturbation can create a giant strongly connected component in an initially subcritical network and destroy it in an initially supercritical one. Importantly, this
behavior is not determined by the transformed in--out-degree distribution alone: in sufficiently asymmetric cases, a reconstruction based only on that distribution can even predict the wrong connectivity phase.

The configuration model provides a useful baseline for interpreting empirical directed networks. However, real-world directed networks may also contain hierarchy, reciprocity, clustering, geometry, and other forms of organization not encoded by the joint degree distribution~\citep{Rodgers2023,Allard2024,BogunaSerrano2025}. We illustrate the resulting framework using a \emph{S. cerevisiae}
transcriptional network distributed by SNAP~\citep{Milo2002,SNAPYeast}
and the San Diego freeway-and-ramp subnetwork extracted from an
OpenStreetMap-derived city network~\citep{Boeing2019,BoeingData,NetzschleuderSD}.
These examples exhibit creation and destruction of strong connectivity,
respectively, treating reversal as a topology-preserving stress test of
how strongly connectivity depends on the observed edge orientations.

\section{Edge-reversal transformation and degree statistics}
\label{sec:model}
Consider a directed graph with joint in--out-degree distribution
$P_0(k_i,k_o)$, where $k_i$ and $k_o$ denote the in- and out-degree of a
vertex, and $k_{\mathrm{tot}} = k_i+k_o$
is its total degree. Each edge is independently reversed with probability
$p$ and retains its original direction with probability $q=1-p$. The
endpoints of every edge remain fixed, so the perturbation changes only
edge directions and leaves the underlying undirected graph unchanged.

Multiple edges and oppositely directed edges between the same pair of
vertices, when present, are treated as distinct labelled edges and are
reversed independently. Allowing such edges is natural for the directed
configuration model and keeps the reversal operation well defined without
additional merging or rejection rules. They are not required for the
degree-level results derived below. In sparse configuration-model limits,
occasional multiple or opposite-direction edges do not affect the leading
percolation threshold, although they may be more relevant in finite,
strongly heterogeneous, or highly reciprocal networks.

Consider now a vertex with initial in-degree $k_i$ and out-degree $k_o$. After
the perturbation, its incoming edges can have two origins: original
incoming edges that were not reversed and original outgoing edges that
were reversed. Let $n$ and $m$ denote the numbers of edges of these two types,
respectively. Then, the new in- and out-degrees are
\begin{equation}
    k_i' = n+m,
    \qquad
    k_o' = k_i+k_o-n-m.
    \label{eq:bin}
\end{equation}
In particular, $k_i'+k_o'=k_i+k_o$, and the total degree of every vertex is exactly conserved. Since edges are reversed independently, $n$ and $m$ are statistically independent and follow binomial distributions with parameters $(k_i,q)$ and $(k_o,p)$, respectively. Therefore, after edges are reversed with probability $p$, the new degree distribution can be written as
\begin{widetext}
\begin{equation}
P_p(k'_i,k'_o)=\sum_{k_i} \sum_{k_o} P_0(k_i,k_o) \sum_{n=0}^{k_i}\sum_{m=0}^{k_o} \binom{k_i}{n} q^n p^{k_i-n}
\binom{k_o}{m} p^m q^{k_o-m}
\delta_{k_i',n+m} \delta_{k'_o,k_i+k_o-n-m}.
\label{eq:ptransformed}
\end{equation}
\end{widetext}
We now define the generating function of the joint in--out-degree distribution after the perturbation by
\begin{equation}
    \Pgf_p(z_i,z_o)
    =
    \sum_{k_i}\sum_{k_o} P_p(k_i,k_o)z_i^{k_i} z_o^{k_o},
    \label{eq:pgf}
\end{equation}
and similarly define $\Pgf_0(z_i,z_o)$ for the original joint in--out-degree distribution. Using Eq.~\eqref{eq:ptransformed}, we see that these two generating functions are related by
\begin{equation}
    \Pgf_p(z_i,z_o)
    =
    \Pgf_0(qz_i+pz_o,\,pz_i+qz_o).
    \label{eq:transform}
\end{equation}
Equation~\eqref{eq:transform} gives the exact change in the joint
in--out-degree distribution under edge-reversal noise. It is a
degree-level identity and does not require a configuration-model,
tree-like, or other random-graph assumption.

Several invariants follow immediately. Since reversal leaves
$k_{\mathrm{tot}}$ unchanged at every vertex, the total-degree
distribution $P_{\mathrm{tot}}(k_{\mathrm{tot}})$ is unchanged. Its generating function,
\begin{equation}
    \widehat{P}_\mathrm{tot}(z_{\mathrm{tot}})
    =
    \Pgf_p(z_{\mathrm{tot}},z_{\mathrm{tot}})=\Pgf_0(z_{\mathrm{tot}},z_{\mathrm{tot}}),
    \label{eq:totalpgf}
\end{equation}
is therefore invariant. The numbers of vertices and edges are also
unchanged, as are all weakly connected components, because the underlying
undirected graph is fixed.

Finally, each directed edge contributes one unit to the total in-degree
and one unit to the total out-degree of the network. Their mean values
are therefore equal and remain invariant under reversal:
\begin{equation}
    \av{k_i}_p
    =
    \av{k_o}_p
    =
    \mu,
    \qquad
    \mu
    =
    \frac{\av{k_{\mathrm{tot}}}}{2}.
    \label{eq:mu}
\end{equation}

\subsection{Iteration of the transformation and stationary orientation statistics}
\label{sec:iteration}
The transformation can be iterated. Two independent reversal rounds with probabilities $p_1$ and $p_2$ reverse an edge relative to its initial orientation exactly when one, but not both, reversals occur. This happens with probability $p_1+p_2-2p_1p_2$. After $n$ rounds with the same $p$, the probability to finally reverse an edge is the probability that during the $n$ rounds the number of partial reversals is an odd number. This is simply the sum for all odd numbers of a binomial distribution with parameters $n$ and $p$, which gives
\begin{equation}
 p_n=\frac{1-(1-2p)^n}{2},\qquad q_n=1-p_n,
 \label{eq:pn}
\end{equation}
and the joint degree generating function becomes
\begin{equation}
 \Pgf_n(z_i,z_o)=\Pgf_0(q_nz_i+p_nz_o,\,p_nz_i+q_nz_o).
 \label{eq:npgf}
\end{equation}
The factor $1-2p$ in Eq.~\eqref{eq:pn} carries the memory of the initial orientation. After $n$ rounds,
this directional memory is reduced by the factor $(1-2p)^n$. For $0<p<1$, we have $|1-2p|<1$, so this memory vanishes as
$n\to\infty$ and $p_n\longrightarrow \frac{1}{2}$.
The limiting joint degree distribution is therefore
\begin{equation}
\begin{gathered}
 \Pgf_{\infty}(z_i,z_o)=\Pgf_0\left(\frac{z_i+z_o}{2},\frac{z_i+z_o}{2}\right),\\
P_{\infty}(k_i,k_o)=
P_{{\mathrm{tot}}}(k_i+k_o)
\binom{k_i+k_o}{k_i}
2^{-(k_i+k_o)}.
 \end{gathered}
 \label{eq:fixed}
\end{equation}
Equation~\eqref{eq:fixed} has a direct interpretation. Conditional on a
vertex having total degree $k_{\mathrm{tot}}=k_i+k_o$, each of its incident edges is
equally likely to point inward or outward in the limiting ensemble.
Hence, the limiting network remains directed, but its edge orientations no
longer retain information about their initial directions and every vertex has equal expected value for its in- and out-degree, while the actual values fluctuate from vertex to vertex. Thus, repeated reversal
converges to the uniform distribution over all possible orientations of
that skeleton. At $p=1/2$, this distribution is reached in a single
round, since every edge is then equally likely to have either
orientation. 

The endpoints are exceptional. At $p=0$ no edge changes
direction, whereas at $p=1$ every edge reverses at every round, producing
a deterministic two-cycle between the original network and its global
reversal. Finally, global reversal does not change the strongly connected
components (SCC). Moreover, reversal with probability $1-p$ is equivalent,
up to a global reversal, to reversal with probability $p$. SCC
statistics are therefore symmetric under
$p\leftrightarrow1-p$. We consequently restrict attention to
$0\leq p\leq1/2$, over which increasing $p$ corresponds to increasing edge 
orientation noise.

\subsection{Exact evolution of second-order degree moments}
\label{sec:moments}
We next consider how reversal noise changes the second-order degree
moments. Define the initial second factorial moments of the in- and
out-degree and the mixed factorial moment as
\begin{equation}
 F_i \equiv \av{k_i(k_i-1)}_0  , \; \;  F_o \equiv \av{k_o(k_o-1)}_0,
 \label{eq:moments}
\end{equation}
and
\begin{equation}
F_{io} \equiv \av{k_i k_o}_0.
\end{equation}
Differentiating the generating-function transformation in
Eq.~\eqref{eq:transform} gives
\begin{align}
 F_i(p) &= q^2F_i+p^2F_o+2pqF_{io},
 \label{eq:Fip}\\
 F_o(p) &= p^2F_i+q^2F_o+2pqF_{io},
 \label{eq:Fop}\\
 F_{io}(p) &= (q^2+p^2)F_{io}+pq(F_i+F_o).
 \label{eq:Fiop}
\end{align}
Thus, the second-order degree moments after one reversal step are
determined exactly by the corresponding moments of the original network.

For the mixed factorial moment, Eq.~\eqref{eq:Fiop} gives
\begin{align}
 F_{io}(p)-F_{io}
& =
 pq\left(F_i+F_o-2F_{io}\right)\nonumber\\
& =
 pq\left[
 \av{(k_i-k_o)^2}_0
 -
 \av{k_{\mathrm{tot}}}_0
 \right].
 \label{eq:signFio}
\end{align}
The sign of the quantity in brackets therefore determines whether
reversal noise increases or decreases the mixed in--out degree moment.

After $n$ rounds with the same reversal probability $0<p<1$, the mixed
factorial moment is
\begin{equation}
 F_{io,n}
 =
 F_{io,\infty}
 +(1-2p)^{2n}
 \left(
 F_{io}-F_{io,\infty}
 \right),
 \label{eq:Fion}
\end{equation}
where
\begin{equation}
F_{io,\infty}
 =
 \frac{\av{k_{\mathrm{tot}}(k_{\mathrm{tot}}-1)}_0}{4}
 =
 \frac{F_i+F_o+2F_{io}}{4}.
 \label{eq:Fiolim}
\end{equation}
Since $(1-2p)^{2n}\to0$ as $n\rightarrow\infty$ for $0<p<1$,
Eq.~\eqref{eq:Fion} shows explicitly that $F_{io,n}$ converges to
$F_{io,\infty}$. At $p=1/2$, the random-orientation limit is reached
after a single reversal step.

The relevance of Eq.~\eqref{eq:signFio} stems from the fact that, for uncorrelated random networks, the percolation transition takes place when
\begin{equation}
\frac{\langle k_i k_o \rangle}{\langle k_i \rangle} >1.
\end{equation}
Therefore, the possibility of increasing or decreasing the mixed in--out degree moment opens up the possibility that percolated networks may be destroyed by noise or that non-percolated ones may percolate. However, as we explain in the next section, the new percolation transition cannot be directly derived from Eq.~\eqref{eq:signFio}.

\section{Percolation under edge reversal with fixed edge pairings}
\label{sec:percolation}

The results in the section above are exact averages over
the edge-reversal process for a fixed initial graph. They describe how
the degree statistics change under reversal noise, but they do not by
themselves determine the existence or size of giant directed components.

Knowing only the new joint in--out-degree distribution after reversal is
not enough to predict the percolation behavior of the actual reversed
network. In an ordinary directed configuration model, following a
randomly chosen directed edge leads to a vertex with probability
proportional to its in-degree. The expected number of outgoing edges
available from the vertex reached in this way is therefore
\begin{equation}
 \frac{\av{k_i k_o}}{\av{k_i}}
 =
 \frac{F_{io}}{\mu}.
 \label{eq:perc0}
\end{equation}
This quantity is the mean branching factor for directed exploration.
The percolation threshold occurs when it reaches unity,
$F_{io}/\mu=1$~\citep{Newman2001}.

One might therefore try to apply Eq.~\eqref{eq:perc0} directly to the
transformed degree distribution. This would amount to constructing a new
directed configuration model from the transformed in- and out-degrees,
with the stubs paired uniformly at random, with branching factor
\begin{equation}
 R_{\mathrm{deg}}(p)
 =
 \frac{F_{io}(p)}{\mu}
 =
 \frac{(q^2+p^2)F_{io}+pq(F_i+F_o)}{\mu}.
 \label{eq:degreeonly}
\end{equation}
Such a construction preserves
the transformed degree statistics, but not the edge pairings of the
original graph. It therefore differs from the reversal process studied
here, in which the same pairs of vertices remain connected and only the
edge directions change.

This distinction matters because the preserved pairings contain
information relevant to directed percolation that is not encoded in the
transformed degree distribution alone. After reversal, a current outgoing
edge can arise in two different ways: it can be an original outgoing edge
that retained its direction, or an original incoming edge that was
reversed. Following these two types of edges reaches vertices through
different original stub types and therefore samples different
excess-degree distributions. This motivates a two-type branching
description, consistent with hidden-stub and correlated-percolation
formalisms~\citep{Soderberg2003,Boguna2005,Allard2015}.

\subsection{Two-type branching matrix}
From this point on, we introduce a directed percolation calculation for the problem of edge reversal with fixed edge pairings studied here. We consider the large $N$ limit of sparse directed configuration models with joint degree
distribution converging to $P_0(k_i,k_o)$. The total numbers of incoming and
outgoing stubs are equal, the stubs are initially matched uniformly at
random, and neighborhoods are locally tree-like, so following
directed edges through the network can be approximated by a branching
process implying that each newly reached vertex generates a number of
outgoing edges that can be followed to new vertices, while repeated
visits to previously reached vertices are negligible at this stage.
A giant directed component appears when this exploration has a positive
probability of continuing indefinitely in the large-network limit. Survival of this branching process then identifies the
appearance of extensive directed components, including the Strongly Connected Component (SCC) of nodes that are mutually reachable following directed paths. This is the standard
setting used in giant-SCC results for directed degree-sequence
models~\citep{CooperFrieze2004,DonderwinkelXie2024}.

The only additional information needed after reversal is whether a
current edge kept its original direction or was reversed. We therefore
distinguish two types of edge in the branching process: retained edges
and reversed edges. To see why the two types must be distinguished, consider following a
current outgoing edge and arriving at its target vertex. If the edge was
retained, then it was originally an incoming edge of the vertex that we
reach. A vertex with original degrees $(k_i,k_o)$ is therefore reached with
probability proportional to its original in-degree, and so distributed as
\begin{equation}
 \frac{k_iP_0(k_i,k_o)}{\mu}.
 \label{eq:retained}
\end{equation}
Once there, its $k_o$ original outgoing edges remain outgoing with
probability $q$, while its remaining $k_i-1$ original incoming edges
become outgoing if they are reversed, with probability $p$.

If instead the edge that we followed was itself reversed, then it was
originally an outgoing edge of the vertex that we reach. Vertices are
therefore sampled with probability
\begin{equation}
 \frac{k_oP_0(k_i,k_o)}{\mu}.
 \label{eq:reversed}
\end{equation}
In this case, one of the $k_o$ original outgoing edges has already been
used to reach the vertex, leaving $k_o-1$ such edges, while all $k_i$
original incoming edges remain available to become outgoing through
reversal.

To construct the mean offspring matrix, we average the numbers of
available outgoing edges using the degree distribution conditioned on
the arrival type. We label retained edges as type~1 and reversed edges
as type~2, and define $M_{ab}(p)$ as the mean number of outgoing edges
of type~$b$ available at a vertex reached through an edge of type~$a$.
Here, outgoing always refers to the direction after the perturbation.

For arrival through a retained edge, the original degrees of the
vertex reached are distributed according to
Eq.~\eqref{eq:retained}. Its $k_o$ original outgoing edges are retained
with probability $q$, while each of its $k_i-1$ remaining original
incoming edges becomes outgoing with probability $p$. Therefore,
\begin{align*}
 M_{11}(p)
 &=
 \sum_{k_i,k_o}
 \frac{k_iP_0(k_i,k_o)}{\mu}\,qk_o
 =
 \frac{qF_{io}}{\mu},
 \\
 M_{12}(p)
 &=
 \sum_{k_i,k_o}
 \frac{k_iP_0(k_i,k_o)}{\mu}\,p(k_i-1)
 =
 \frac{pF_i}{\mu}.
\end{align*}
The subtraction of one in the second expression excludes the original
incoming edge through which the vertex was reached.

For arrival through a reversed edge, the corresponding degree
distribution is Eq.~\eqref{eq:reversed}. In this case, the arrival edge
was originally outgoing, leaving $k_o-1$ original outgoing edges that
can retain their direction, while all $k_i$ original incoming edges
can become outgoing through reversal. Thus,
\begin{align*}
 M_{21}(p)
 &=
 \sum_{k_i,k_o}
 \frac{k_oP_0(k_i,k_o)}{\mu}\,q(k_o-1)
 =
 \frac{qF_o}{\mu},
 \\
 M_{22}(p)
 &=
 \sum_{k_i,k_o}
 \frac{k_oP_0(k_i,k_o)}{\mu}\,pk_i
 =
 \frac{pF_{io}}{\mu}.
\end{align*}

Combining these four conditional averages gives the mean offspring
matrix
\begin{equation}
 M(p)=\frac{1}{\mu}
 \begin{pmatrix}
 qF_{io} & pF_i\\
 qF_o    & pF_{io}
 \end{pmatrix}.
 \label{eq:matrix}
\end{equation}
The rows specify whether the current vertex was reached through a
retained or a reversed edge, respectively, and the columns specify
whether the next edge followed is retained or reversed.

The same construction can be carried out backward, by following edges
against their direction. This interchanges $F_i$ and $F_o$, but the
resulting matrix has the same eigenvalues. Consequently, forward and
backward exploration have the same growth threshold. The largest
eigenvalue of $M(p)$ is
\begin{equation}
 \Lambda(p)=
 \frac{
 F_{io}
 +\sqrt{(1-2p)^2F_{io}^2+4pqF_iF_o}
 }{2\mu}.
 \label{eq:lambda}
\end{equation}
In the Appendix~\ref{app:components}, we show that this eigenvalue is the effective branching factor of the two-type
process. In the configuration-model regime described above, directed
exploration becomes extensive when
\[
 \Lambda(p)>1.
\]
The same condition governs the appearance of giant in- and
out-components and, when both are present, of a giant strongly connected
component. The corresponding component-size equations are derived in
Appendix~\ref{app:components}.

At $p=0$, no edge is reversed, so the two-type description reduces to
the usual directed configuration-model result,
\begin{equation}
 \Lambda(0)=\frac{F_{io}}{\mu}.
  \label{eq:endpoints0}
\end{equation}
At $p=1/2$, each edge direction is randomized with equal probability,
and
\begin{equation}
 \Lambda(1/2)
 =
 \frac{F_{io}+\sqrt{F_iF_o}}{2\mu}.
 \label{eq:endpoints}
\end{equation}

For comparison, let $R_{\mathrm{deg}}(p)$ denote the branching factor
obtained by discarding the original edge pairings and constructing a new
directed configuration model from the transformed degree distribution
alone. At $p=1/2$, the difference between this degree-only prediction and
the branching factor of the actual reversal process is
\begin{equation}
 R_{\mathrm{deg}}(1/2)-\Lambda(1/2)
 =
 \frac{(\sqrt{F_i}-\sqrt{F_o})^2}{4\mu}
 \geq 0.
 \label{eq:endpointdifference}
\end{equation}
Thus, even when the edge directions have been completely randomized, a
network obtained by reversing directions while preserving the original
edge pairings is generally not equivalent to a new configuration-model
network constructed only from the resulting degree distribution.

\subsection{Creation, destruction, and repeated reversal}
On the interval $0\leq p\leq1/2$, the quantity
$pq=p(1-p)$ increases monotonically from $0$ to $1/4$, while
$(1-2p)^2$ decreases from $1$ to $0$. It follows from
Eq.~\eqref{eq:lambda} that the branching factor $\Lambda(p)$ is itself
monotonic:
\begin{align}
 &\Lambda(p)\ \text{increases if}\ F_iF_o>F_{io}^2,\nonumber\\
 &\Lambda(p)\ \text{decreases if}\ F_iF_o<F_{io}^2,\nonumber\\
 &\Lambda(p)\ \text{remains constant if}\ F_iF_o=F_{io}^2.
\end{align}

Because $\Lambda(p)$ is monotonic on $0\leq p\leq1/2$, the possibility
of a noise-induced transition can be determined entirely from its values
at the two endpoints. Using Eq.(\ref{eq:endpoints0}) and Eq.(\ref{eq:endpoints}), a noise-induced creation of a giant strongly connected component occurs
when
\begin{equation}
 \frac{F_{io}}{\mu}
 <
 1
 <
 \frac{F_{io}+\sqrt{F_iF_o}}{2\mu},
 \label{eq:createcondition}
\end{equation}
so that the network is subcritical at $p=0$ but becomes supercritical
as the directions are randomized. Conversely, a destruction transition
occurs when
\begin{equation}
 \frac{F_{io}}{\mu}
 >
 1
 >
 \frac{F_{io}+\sqrt{F_iF_o}}{2\mu},
 \label{eq:destroycondition}
\end{equation}
so that a giant strongly connected component present initially is lost
as the reversal probability increases. If both endpoint values are on
the same side of one, no percolation transition occurs on
$0\leq p\leq1/2$.

When the endpoint values lie strictly on opposite sides of one,
monotonicity guarantees a unique transition probability
$0<p_c<1/2$. At the transition,
\[
 \Lambda(p_c)=1.
\]
Defining $\eta_c=p_c(1-p_c)$, after some algebra we obtain
\begin{equation}
 \eta_c
 =
 \frac{\mu(F_{io}-\mu)}
 {F_{io}^2-F_iF_o},
 \qquad
 p_c
 =
 \frac{1-\sqrt{1-4\eta_c}}{2}.
 \label{eq:pc}
\end{equation}
The minus sign in the expression for $p_c$ selects the solution in the
interval $0<p_c<1/2$. Finally, requiring
$\Lambda(0)$ and $\Lambda(1/2)$ to lie on opposite sides of one ensures
that Eq.~\eqref{eq:pc} corresponds to the physical percolation
transition.

The same framework also applies to repeated reversal. After $n$ rounds
with the same reversal probability $p$, an edge has the effective
reversal probability $p_n$ given by Eq.~\eqref{eq:pn}. Substituting
$p_n$ into Eq.~\eqref{eq:lambda} gives
\begin{equation}
 \Lambda_n
 =
 \frac{
 F_{io}
 +
 \sqrt{
 F_iF_o
 +(1-2p)^{2n}
 \left(F_{io}^2-F_iF_o\right)
 }
 }{2\mu}.
 \label{eq:lambdan}
\end{equation}
For $0<p<1$, $(1-2p)^{2n}\to0$ as $n\rightarrow\infty$, and therefore
\begin{equation}
 \Lambda_n
 \longrightarrow
 \frac{F_{io}+\sqrt{F_iF_o}}{2\mu}
 =
 \Lambda(1/2).
\end{equation}
If $F_iF_o>F_{io}^2$, this limiting value is approached from below; if
$F_iF_o<F_{io}^2$, it is approached from above. When
$F_iF_o=F_{io}^2$, the branching factor is unchanged by reversal.

\section{Examples}
\label{sec:examples}

\subsection{Creation from sources and sinks in a bipartite directed network}
\begin{figure}[t]
 \centering
 \includegraphics[width=\columnwidth]{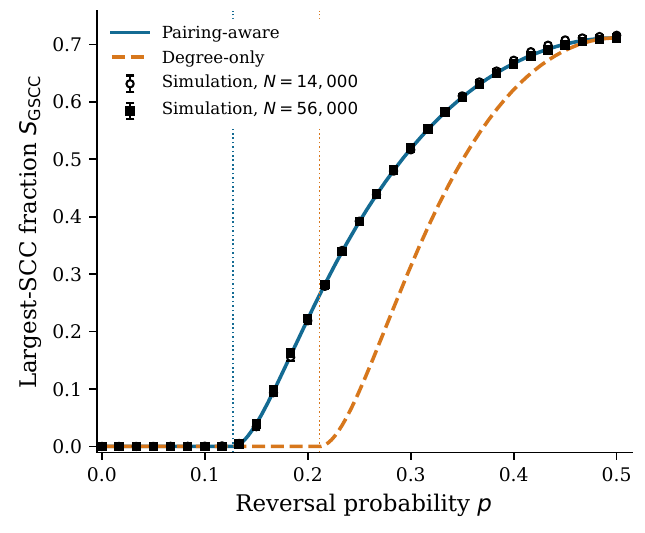}
 \caption{Creation of a giant strongly connected component in the $k=4$ bipartite ensemble. The solid curve is the solution of the two-type component equations in Appendix~\ref{app:components}. Symbols show the mean largest-SCC fraction from synthetic configuration-model simulations. Blue and orange dotted vertical lines mark the two-type and degree-only threshold predictions, respectively. The orange dashed curve is the degree-only prediction. Open circles and filled squares show the smaller and larger simulated sizes, respectively; error bars indicate one standard error over 12 realizations.}
 \label{fig:creation}
\end{figure}
Consider an initially acyclic configuration-model ensemble in
which half of the vertices are pure sources, with degrees $(0,k)$, and
the other half are pure sinks, with degrees $(k,0)$. Thus every edge in
the original bipartite network points from a source to a sink. A directed path
therefore stops as soon as it reaches a sink, so the initial network
cannot contain a giant strongly connected component.

The joint degree generating function is
\begin{align}
 \Pgf_0(x,y)&=\frac{x^k+y^k}{2},\nonumber\\
 \mu&=\frac{k}{2}, &
 F_i=F_o&=\frac{k(k-1)}{2}, &
 F_{io}&=0.
 \label{eq:source}
\end{align}
The vanishing mixed moment $F_{io}=0$ reflects the complete separation
between sources and sinks: no vertex has both incoming and outgoing
edges initially.

Substituting these moments into Eq.~\eqref{eq:lambda} gives
\begin{equation}
 \Lambda(p)=(k-1)\sqrt{pq}.
 \label{eq:sourcelambda}
\end{equation}
At $p=0$, $\Lambda(0)=0$, consistent with the absence of sustained
directed paths. As edges are reversed, however, some of the original
source-to-sink edges become sink-to-source edges. Sinks can then acquire
outgoing edges and sources can acquire incoming edges, allowing directed
paths to alternate between the two vertex classes and creating return
routes that were absent in the original orientation.

A giant strongly connected component appears when
$\Lambda(p)$ crosses one. Solving $\Lambda(p_c)=1$ gives
\begin{equation}
 p_c
 =
 \frac{1-\sqrt{1-4/(k-1)^2}}{2},
 \label{eq:sourcepc}
\end{equation}
where the root in the interval $0\leq p_c\leq1/2$ has been chosen.
Such a transition is possible only for $k\geq4$, because at complete
direction randomization
\[
 \Lambda(1/2)=\frac{k-1}{2},
\]
which exceeds one only when $k>3$.

For the simplest case, $k=4$, $p_c\simeq0.127322$. By contrast, a degree-only reconstruction based on the transformed
degrees gives $R_{\mathrm{deg}}(p)=6pq$, and would place the transition at
$p\simeq0.211325$. The difference arises because the degree-only
reconstruction discards the original source--sink edge pairings, whereas
the actual reversal process preserves them.

Importantly, the transition is produced without adding or removing any
edges. Reversal changes only their directions, but this is sufficient to
create directed return paths on the same underlying bipartite network
and thereby generate a giant strongly connected component. Figure~\ref{fig:creation} shows numerical simulations of this toy example confirming our predictions.

\subsection{Destruction in an initially balanced directed ensemble}
\begin{figure}[t]
 \centering
 \includegraphics[width=\columnwidth]{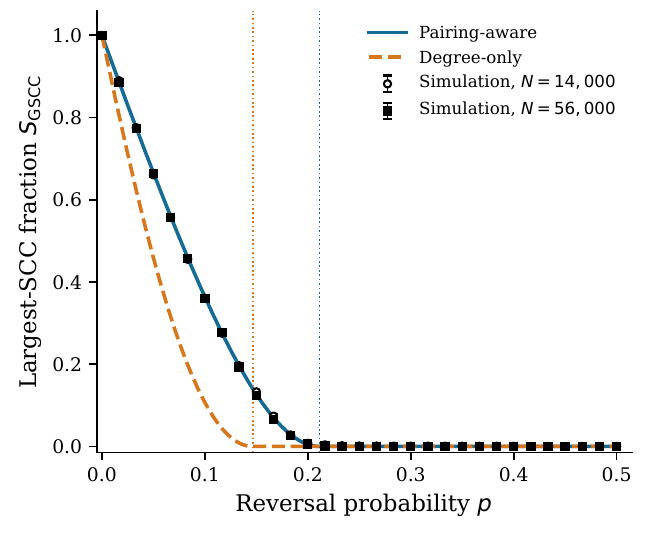}
 \caption{Destruction of the giant strongly connected component for Eq.~\eqref{eq:balanced}. The branching prediction is $p_c\simeq0.211325$, rather than the degree-only value $0.146447$. The blue solid and orange dashed curves show the pairing-aware and degree-only predictions; matching dotted vertical lines mark their thresholds. Open circles and filled squares show the smaller and larger simulated sizes, respectively, with one-standard-error bars over 12 realizations. Finite graphs retain small SCCs in the subcritical regime, while the extensive fraction vanishes in the random-graph limit.}
 \label{fig:destruction}
\end{figure}

Consider an initially balanced directed configuration-model ensemble with
joint degree generating function
\begin{equation}
 \Pgf_0(x,y)=\frac67xy+\frac17x^2y^2.
 \label{eq:balanced}
\end{equation}
Thus, a fraction $6/7$ of the vertices have one incoming and one
outgoing edge, $(k_i,k_o)=(1,1)$, while the remaining fraction $1/7$
have two incoming and two outgoing edges, $(k_i,k_o)=(2,2)$. Every
vertex therefore has equal in- and out-degree in the original network.

The corresponding moments are
\begin{equation}
 \mu=\frac87,\qquad
 F_i=F_o=\frac27,\qquad
 F_{io}=\frac{10}{7}.
 \label{eq:balancedmoments}
\end{equation}
At $p=0$, the branching factor is
\[
 \Lambda(0)=\frac{F_{io}}{\mu}=\frac54>1,
\]
so the initial network is in the supercritical regime and supports a
giant strongly connected component.

Edge reversal has the opposite effect from the previous bipartite
example. Although the mean in- and out-degrees remain unchanged,
individual vertices need no longer remain balanced. For example, a
vertex that initially has one incoming and one outgoing edge can, after
reversal, end up with two incoming edges or two outgoing edges. Such
local imbalances reduce the ability of directed paths to continue in
both directions through the network.

Substituting the moments in Eq.~\eqref{eq:lambda} for the effective branching factor of the two type process gives
\begin{equation}
 \Lambda(p)
 =
 \frac{5+\sqrt{25-96pq}}{8}.
 \label{eq:balancedlambda}
\end{equation}
Since $F_iF_o<F_{io}^2$, this quantity decreases monotonically as $p$
increases from $0$ to $1/2$. At complete randomization of the edge
directions,
\[
 \Lambda(1/2)=\frac34<1.
\]
The network therefore passes from the supercritical to the subcritical
regime as the reversal probability increases.

The destruction threshold is obtained from $\Lambda(p_c)=1$, giving
\begin{equation}
 p_c
 =
 \frac{1-1/\sqrt3}{2}
 \simeq0.211325.
 \label{eq:balancedpc}
\end{equation}
Thus, once approximately $21\%$ of the edges are independently reversed,
the branching process becomes subcritical in the large-network limit.

For comparison, reconstructing a new directed configuration model from
the transformed degree distribution alone gives
\[
 R_{\mathrm{deg}}(p)
 =
 \frac54-2pq,
\]
which crosses one at
\[
 p\simeq0.146447.
\]
The degree-only reconstruction therefore predicts the loss of the giant
strongly connected component substantially earlier than the
pairing-preserving reversal calculation. The difference arises because
the actual reversal process keeps the original edge pairings, whereas
the degree-only construction retains only the transformed degree
statistics. Figure~\ref{fig:destruction} shows numerical simulations, confirming again our predictions.

\subsection{Degree-only reconstruction can predict the wrong percolation phase}
The previous two examples have $F_i=F_o$. In that symmetric case, the
degree-only and pairing-preserving branching criteria become equal at
$p=1/2$, although they can differ for intermediate reversal
probabilities. To show that the distinction can be qualitative rather
than merely quantitative, consider the degree distribution with characteristic function
\begin{equation}
 \Pgf_0(x,y)
 =
 \frac14x^4y+\frac58y+\frac18xy^2.
 \label{eq:asym}
\end{equation}
This distribution contains three types of vertices: a fraction $1/4$
have degrees $(k_i,k_o)=(4,1)$, a fraction $5/8$ have degrees $(0,1)$,
and a fraction $1/8$ have degrees $(1,2)$. The network is therefore
strongly asymmetric in its in- and out-degree statistics, even though
the mean in- and out-degrees remain equal, as required for a directed
network.

The corresponding moments are
\begin{equation}
 \mu=\frac98,\qquad
 F_i=3,\qquad
 F_o=\frac14,\qquad
 F_{io}=\frac54.
 \label{eq:asymmom}
\end{equation}
Two combinations of these moments control two different responses to
reversal noise:
\begin{equation}
 F_i+F_o-2F_{io}
 =
 \frac34>0,
 \qquad
 F_iF_o-F_{io}^2
 =
 -\frac{13}{16}<0.
 \label{eq:oppositesigns}
\end{equation}
The first quantity determines the change of the mixed factorial moment
$F_{io}(p)$, while the second determines the change of the true
branching factor $\Lambda(p)$. Their opposite signs therefore imply that
these two quantities move in opposite directions as $p$ increases.

Indeed, the degree-only reconstruction gives
\begin{equation}
 R_{\mathrm{deg}}(p)
 =
 \frac{10}{9}+\frac23pq,
 \label{eq:asymdeg}
\end{equation}
which increases with reversal noise and remains larger than one for the
entire interval $0\leq p\leq1/2$. If one considered only the transformed
degree distribution, one would therefore conclude that the network
remains in the supercritical regime.

The pairing-preserving branching calculation gives a different result:
\begin{equation}
 \Lambda(p)
 =
 \frac{5+\sqrt{25-52pq}}{9}.
 \label{eq:asymlambda}
\end{equation}
Because $F_iF_o<F_{io}^2$, $\Lambda(p)$ decreases monotonically with
$p$. Initially,
\[
 \Lambda(0)=\frac{10}{9}>1,
\]
so the network is supercritical. As the edge directions are randomized,
however, the branching factor falls and eventually crosses the
percolation threshold. Solving $\Lambda(p_c)=1$ gives
\begin{equation}
 p_c
 =
 \frac{1-2/\sqrt{13}}{2}
 \simeq0.222650.
 \label{eq:asympc}
\end{equation}
Thus the giant strongly connected component is predicted to disappear
once the reversal probability exceeds approximately $0.223$.

The contrast is especially clear at complete direction randomization,
$p=1/2$:
\begin{align}
 \Lambda(1/2)
 &=
 \frac{5+2\sqrt3}{9}
 \simeq0.94046<1,
 \nonumber\\
 R_{\mathrm{deg}}(1/2)
 &=
 \frac{23}{18}>1.
 \label{eq:asymendpoint}
\end{align}
The two descriptions therefore predict different connectivity phases:
the degree-only reconstruction remains supercritical, whereas the
pairing-preserving calculation is subcritical.

\begin{figure}[t]
 \centering
 \includegraphics[width=\columnwidth]{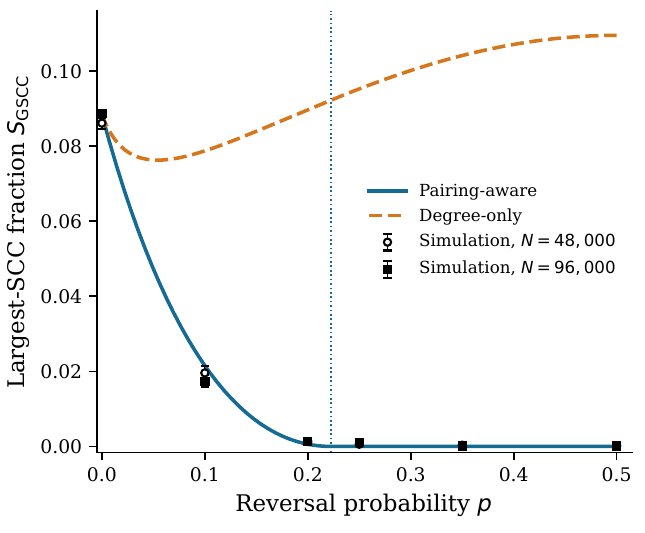}
 \caption{Asymmetric destruction example, Eq.~\eqref{eq:asym}. The two-type theory predicts the loss of the extensive SCC at $p_c\simeq0.22265$, while the degree-only comparison remains supercritical for all $0\leq p\leq1/2$. The blue solid and orange dashed curves show the pairing-aware and degree-only predictions, respectively; the blue dotted line marks the pairing-aware threshold. Open circles and filled squares show the smaller and larger synthetic configuration-model sizes, respectively, with one-standard-error bars over 12 realizations. Sampling near the transition is limited and the figure is intended as a consistency check rather than a finite-size scaling study.}
 \label{fig:asym}
\end{figure}

This example clearly shows why the transformed joint degree distribution alone
is not sufficient to determine percolation under edge reversal. The
quantity $F_{io}(p)/\mu$, which would be the branching factor of a newly
rewired configuration model with the transformed degrees, actually
increases with $p$. Nevertheless, the true branching factor of the
edge-reversal process decreases because the original edge pairings are
preserved and retained and reversed edges sample those pairings
differently. Figure~\ref{fig:asym} shows simulation results confirming these predictions.

\section{Applications to real directed networks}
\label{sec:applications}

For an empirical directed network, edge reversal can be used as a
sensitivity test of how strongly connectivity depends on the observed
edge directions. The network
connections are kept fixed, each edge is independently reversed with
probability $p$, and the resulting response is measured through
\begin{equation}
 S_{\GSCC}(p),\qquad
 S_{\GIN}(p),\qquad
 S_{\GOUT}(p).
 \label{eq:responsecurve}
\end{equation}
These curves show how mutual reachability changes as the edge
orientations are progressively randomized.

The empirical response can be compared with two
configuration-model baselines. The first is the pairing-aware two-type
model developed above, which accounts for the different roles of
retained and reversed edges. The second is the degree-only
reconstruction, in which a new directed configuration model is built
from the transformed degree distribution $P_p(k_i,k_o)$ and the original
edge pairings are discarded. Differences between the empirical response
and either baseline indicate structural information not contained in
that particular random-graph description, such as reciprocity, clustering, geometry, motifs,
and degree correlations~\citep{Rodgers2023,Allard2024,BogunaSerrano2025}.

The same procedure can also be used when edge directions are
uncertain rather than deliberately perturbed. A uniform $p$ provides a
simple reference test, while edge-specific reversal probabilities can be
used when confidence in the inferred directions varies from edge to
edge. 

\subsection{Creation of strong connectivity in a \emph{S. cerevisiae} transcriptional regulation network}
We illustrate this procedure using the \emph{S. cerevisiae}
transcriptional regulation network distributed by SNAP of
Milo \emph{et al.}~\citep{Milo2002,SNAPYeast}. The network contains 690 vertices and 1,094 directed edges,
including 12 self-regulatory loops. Removing those loops leaves $N=690, E=1082$, 
with no isolated vertices. The analysis uses only directed topology;
regulatory signs do not enter the reversal process. The largest weakly
connected component contains 664 vertices, whereas the observed largest
strongly connected component contains only three:
\begin{equation}
 S_{\GSCC}(0)=\frac{3}{690}\simeq0.00435.
\end{equation}
Thus the observed orientation supports very little mutual reachability,
despite an almost network-spanning weakly connected component.

The loop-free network has mean degree and factorial moments
\begin{align}
 \mu&=1.568116, & F_{io}&=0.565217,\nonumber\\
 F_i&=34.730435, & F_o&=2.802899.
\end{align}
The pairing-aware branching factor consequently satisfies
\begin{equation}
 \Lambda(0)=0.36044<1<\Lambda(1/2)=3.32616.
\end{equation}
Unlike a case in which reversal merely enlarges an already-supercritical
core, this degree sequence satisfies the creation criterion. Solving
$\Lambda(p_c)=1$ gives $p_c\simeq0.01648$, corresponding to about 17.8 reversed edges on average. The degree-only
reconstruction instead predicts
$
 p_c^{\mathrm{deg}}\simeq0.02835.
$
These are configuration-model thresholds inferred from the empirical
degree sequence, rather than sharp transition points of this finite
fixed network.

\begin{figure}[t]
 \includegraphics[width=\columnwidth]{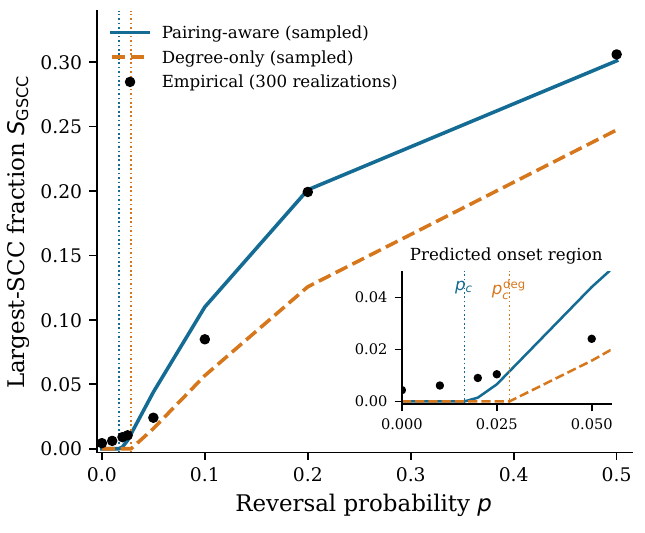}
 \caption{Creation of strong connectivity in the \emph{S. cerevisiae}
 transcriptional network under edge-reversal noise. Symbols show the
 mean largest-SCC fraction on the fixed empirical network over 300
 reversal realizations. The solid and dashed lines connect the reported
 sampled predictions of the pairing-aware two-type and degree-only
 configuration-model baselines, respectively; the connecting segments
 are guides to the eye. Vertical lines mark their predicted thresholds
 $p_c\simeq0.01648$ and $p_c^{\mathrm{deg}}\simeq0.02835$.
 The inset resolves the small-$p$ region. The error bar at $p=1/2$ is
 the reported standard error; uncertainties at the other displayed
 points were not retained in the available numerical summary.}
 \label{fig:yeast}
\end{figure}
Figure~\ref{fig:yeast} compares the empirical reversal response with the
two configuration-model baselines. Each empirical point averages 300
independent reversal realizations. The mean largest-SCC fraction increases to $S_{\GSCC}(0.05)\approx 0.02$ at $p=0.05$, to $S_{\GSCC}(0.5)\approx 0.3$ at $p=0.5$. Randomizing the directions therefore produces a strongly connected core containing about 31\% of the network, starting from less than half a percent.

The pairing-aware baseline predicts $S_{\GSCC}(0.2) \approx 0.2$ and
$S_{\GSCC}(1/2)\approx 0.3$, close to the corresponding empirical values.
The degree-only reconstruction gives $0.12571$ and $0.24740$,
respectively. Retaining the distinction between original and reversed
arrivals therefore improves the description of both the predicted
creation threshold and the upper part of the response curve. Agreement
is less close at intermediate noise: at $p=0.05$ and $p=0.1$, the
two-type baseline gives $0.04399$ and $0.11008$, respectively, slightly larger than the empirical values.

Gene-regulatory networks provide a natural setting for this type of
directionality test because their interpretation depends strongly on
which genes regulate which others, and directed loops can be sensitive
to reversal of even a small fraction of edges~\citep{Itzhack2013}.
Uniform reversal probability remains a deliberately simple
perturbation: real uncertainty in regulatory directions may be
heterogeneous, correlated, or absent for particular interaction types.
Finally, changes in SCC structure should not by themselves be
interpreted as evidence for changes in oscillatory behavior, dynamical
stability, regulatory strength, or biological function.

\subsection{Destruction of strong connectivity in the San Diego freeway network}
\label{sec:sandiego}
A complementary empirical example is provided by the San Diego
freeway-and-ramp subnetwork~\citep{Boeing2019,BoeingData}. We use the
simplified directed multigraph from~\citep{NetzschleuderSD}. We retain edges whose highway classifications
consist exclusively of \texttt{motorway} and/or \texttt{motorway\_link},
together with all their incident vertices. Original edge directions,
parallel edges, and reciprocal arcs are retained; self-loops are
excluded. No restriction to the largest connected component is made.
This construction selects a road class within the supplied city
boundary, rather than the complete urban street network.

The resulting graph contains
$
 N=1608, E=1932,
$
and its largest SCC initially contains 668 vertices, giving
$S_{\GSCC}(0)=668/1608\simeq0.41542$. Its degree moments are
\begin{align}
 \mu&=1.201493, & F_{io}&=1.476990,\nonumber\\
 F_i&=0.730100, & F_o&=0.696517.
\end{align}
Substitution into Eq.~\eqref{eq:lambda} yields
\begin{equation}
 \Lambda(0)=1.22930>1>\Lambda(1/2)=0.91141.
 \label{eq:sandiego_endpoints}
\end{equation}
The empirical degree sequence therefore satisfies the destruction
criterion, complementing the creation example above. The pairing-aware
configuration-model branching factor crosses one at
$
 p_c\simeq0.27165.
$
This value is the threshold of the configuration-model baseline,
not an independently measured critical point of the finite road graph.

\begin{figure}[t]
 \includegraphics[width=\columnwidth]{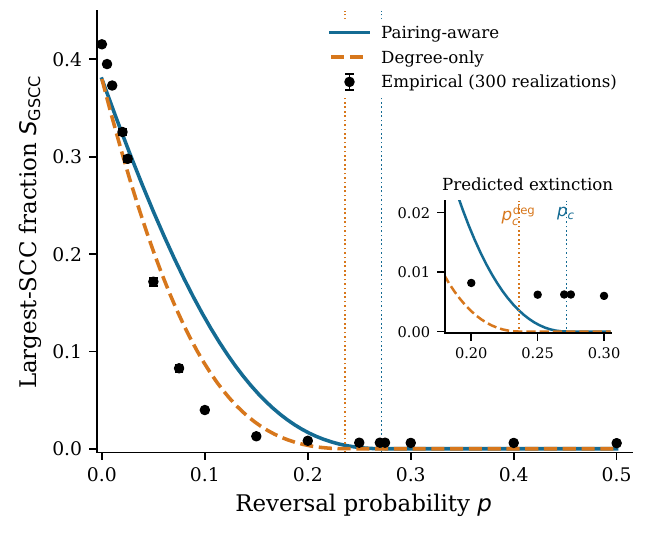}
 \caption{Destruction of strong connectivity in the San Diego
 freeway-and-ramp subnetwork. Mean largest-SCC fraction on the fixed
 empirical graph over 300 independent reversal realizations; error bars
 indicate one standard error and connecting segments are guides to the
 eye.}
 \label{fig:sandiego}
\end{figure}

For each of 300 independent realizations, we assign independent uniform
random variates to the labelled edges and reverse an edge whenever its
variate is below $p$. Figure~\ref{fig:sandiego} shows the resulting mean largest-SCC
fraction. It decreases to $0.17157$ at $p=0.05$, $0.03975$ at $p=0.1$,
and $0.00817$ at $p=0.2$. At fair random orientation,
\begin{equation}
 S_{\GSCC}(1/2)=0.00573\pm0.00013,
\end{equation}
where the uncertainty is the standard error across realizations.
Thus an initial strongly connected core containing about 42\% of the
vertices is reduced to a largest SCC containing about nine vertices on
average.

The endpoint criterion and the empirical reversal response agree on
destruction, but the configuration-model threshold does not accurately
locate the observed decline: the empirical largest SCC is already small
well below $p_c$. Spatial organization and correlations beyond the
single-vertex degree distribution are not represented by this baseline.
The finite-network experiment demonstrates loss of a large strongly
connected core, rather than a finite-size scaling verification of a
critical point. Reversal here is a topological sensitivity test; it does
not model traffic capacities, congestion, travel times, or turn
restrictions.

\section{Discussion}
\label{sec:discussion}
Independent edge reversal changes the directions of connections while
preserving which vertices they join. We obtain an exact description of
how this process changes the incoming and outgoing degrees of each
vertex while leaving its total degree unchanged. Repeated reversal
gradually erases the initial directional organization, making each edge
equally likely to point either way. These local changes alter the
directed paths through the network and can therefore reshape its
large-scale connectivity.

Understanding this global response requires keeping track of how the
connections were originally arranged. Networks with the same incoming
and outgoing degree distributions can respond differently because
reversal preserves the original connections. Our theory captures this
information by distinguishing paths that arrive along retained edges
from those that arrive along reversed edges. This distinction leads
to explicit conditions under which reversal creates or destroys a
giant strongly connected component.

The synthetic examples show why this distinction matters. Reversal
can create strong connectivity in an initially acyclic source--sink
network or destroy it in an initially balanced directed ensemble.
In the asymmetric example, the transformed mixed moment increases
while the pairing-aware branching factor decreases. A degree-only
reconstruction consequently predicts a supercritical network even
where the pairing-preserving ensemble is subcritical. The transformed
degree statistics therefore need not identify even the correct
percolation phase.

The empirical examples illustrate both the usefulness and the limits
of the theory. In the \emph{S. cerevisiae} regulatory network, the
degree moments satisfy the creation criterion, and reversal expands
the largest SCC from approximately $0.435\%$ to $30.6\%$ of the
vertices. The pairing-aware prediction closely describes the response
at larger reversal probabilities, whereas the degree-only
reconstruction underestimates the resulting core. In the San Diego
freeway-and-ramp subnetwork, the moments instead satisfy
$\Lambda(0)>1>\Lambda(1/2)$, and the largest SCC decreases from
approximately $41.5\%$ to $0.573\%$ of the vertices. The endpoint
criterion agrees with the observed destruction, although the empirical
decline occurs substantially earlier than the predicted
configuration-model threshold.

The scope of these results depends on the level of description.
The degree transformation applies to arbitrary directed networks,
whereas the percolation predictions rely on the locally tree-like
configuration-model ensemble. Geometry, clustering, reciprocity,
and correlations between adjacent vertices can therefore produce
departures from the predicted response. A small nonzero largest SCC
in a finite graph also remains compatible with the absence of an
extensive component in the asymptotic theory. The empirical examples
demonstrate creation and destruction of large strongly connected
cores, but do not establish the predicted critical points through
finite-size scaling.

Directional uncertainty provides a further motivation for this
analysis. In regulatory networks inferred from gene-expression data,
a statistical association may be easier to establish than which gene
regulates the other. In causal networks inferred from observational
data, distinct edge orientations can be compatible with the same
observed dependencies. Brain networks reconstructed from diffusion
MRI present a related limitation: the measurements do not distinguish
afferent from efferent pathways and therefore generally yield
undirected connections. These settings motivate testing how
connectivity conclusions depend on assumed orientations. Independent
reversal provides a simple reference perturbation, although realistic
uncertainty may be heterogeneous or correlated, and additional
constraints, such as acyclicity in causal models, require a different
orientation-sampling procedure. Our empirical experiments test
directional sensitivity rather than estimate errors in the recorded
edge directions.

Edge reversal provides a controlled probe of how strongly mutual
reachability depends on direction when the underlying connections
remain fixed. The pairing-aware theory supplies an explicit baseline
for this experiment: it identifies when orientation changes alone
can create or destroy extensive strong connectivity, while departures
from its predictions reveal the need to account for network structure
beyond the original degree distribution.

More broadly, these results show that the same set of connections can
support markedly different patterns of mutual accessibility depending
on its orientation. Directional disorder can either enable or suppress
large-scale connectivity, so randomizing directions need not simply
degrade an existing network. This dual role makes directionality an
independent dimension of network organization and motivates studying
orientation changes alongside the addition, removal, or rewiring of
connections.

\section*{Data and code}
The empirical network data were obtained from publicly available
sources: SNAP for the \emph{S. cerevisiae} transcriptional
network~\citep{Milo2002,SNAPYeast}, and the OpenStreetMap-derived
collection distributed through Netzschleuder for the San Diego
network~\citep{Boeing2019,BoeingData,NetzschleuderSD}.
All data generated for the analyses and the code used for network
preprocessing, numerical simulations, analytical verification, and
figure generation are available from the authors upon request.

\section*{Acknowledgments}
The authors acknowledge support from Grant No. PID2022-137505NB-C22 funded by MCIN/AEI/10.13039/501100011033 and by ERDF/EU and Grant No. PID2025-169261NB-C21 funded by MICIU/AEI/10.13039/501100011033.

\appendix

\section{Giant-component equations}
\label{app:components}

We derive the sizes of the giant directed components using the
finite-component generating-function approach of
Ref.~\citep{Boguna2005}. We consider the sparse, locally tree-like
configuration-model regime described in the main text, with original
joint degree distribution $P_0(k_i,k_o)$. Each labelled edge has a
single direction and is independently reversed with probability $p$,
while its endpoints remain fixed. There are therefore no undirected
or bidirectional edges in the branching description.

The configuration-model assumption applies to the original network.
Its incoming and outgoing stubs are matched uniformly at random, so
the original transition probabilities are
\begin{align}
 P_o^{(0)}(\mathbf{k}'\mid\mathbf{k})
 &=
 \frac{k_i'P_0(\mathbf{k}')}{\mu},
 \nonumber\\
 P_i^{(0)}(\mathbf{k}'\mid\mathbf{k})
 &=
 \frac{k_o'P_0(\mathbf{k}')}{\mu},
 \label{eq:B_original_transitions}
\end{align}
where $\mathbf{k}=(k_i,k_o)$ denotes the original degrees and
$\mu=\av{k_i}_0=\av{k_o}_0$. Correlations between the in- and
out-degree of the same vertex are allowed. After reversal, we retain
the distinction between edges that kept their direction and edges
that reversed, since these two types sample different original
stub distributions.

\subsection{Finite-component distributions}

Let $G_o(s_o\mid\mathbf{k})$ denote the probability that a vertex with
original degrees $\mathbf{k}$ has a finite out-component containing
$s_o$ vertices, including the vertex itself. Similarly,
$G_i(s_i\mid\mathbf{k})$ is the probability that its finite
in-component contains $s_i$ vertices. We also define $G(s_i,s_o\mid\mathbf{k})$ as the probability
that both components are finite and have sizes $s_i$ and $s_o$,
respectively. Their generating functions are
\begin{align}
 \widehat G_o(z_o\mid\mathbf{k})
 &=
 \sum_{s_o\geq1}G_o(s_o\mid\mathbf{k})z_o^{s_o},
 \nonumber\\
 \widehat G_i(z_i\mid\mathbf{k})
 &=
 \sum_{s_i\geq1}G_i(s_i\mid\mathbf{k})z_i^{s_i},
 \nonumber\\
 \widehat G(z_i,z_o\mid\mathbf{k})
 &=
 \sum_{s_i,s_o\geq1}G(s_i,s_o\mid\mathbf{k})z_i^{s_i} z_o^{s_o}.
 \label{eq:B_component_generators}
\end{align}
The corresponding unconditional generating functions $\widehat G_o(z_o)$, $\widehat G_i(z_i)$, and $\widehat G(z_i,z_o)$ are obtained
by averaging over $P_0(\mathbf{k})$. All distributions include the
average over the independent edge reversals.

The sums run only over finite component
sizes. Their total probability can therefore be smaller than one
above the percolation threshold. In particular,
$\widehat G_o(1)$ is the probability that the out-component is finite,
whereas $\widehat G(1,1)$ is the probability that both components are
finite. Consequently, $\widehat G(1,z)$ is generally different from
$\widehat G_o(z)$ above the threshold. Values of the generating
functions at unity are understood as limits from below.

With the convention that $\GIN$ and $\GOUT$ both include the
$\GSCC$, their relative sizes are
\begin{equation}
 S_{\GIN}=1-\widehat G_o(1),
 \qquad
 S_{\GOUT}=1-\widehat G_i(1).
 \label{eq:B_giant_marginals}
\end{equation}
A vertex belongs to the $\GIN$ when its out-component is extensive,
and to the $\GOUT$ when its in-component is extensive. Membership
in the $\GSCC$ requires both conditions. Inclusion--exclusion
therefore gives
\begin{equation}
 S_{\GSCC}
 =
 1-\widehat G_o(1)-\widehat G_i(1)+\widehat G(1,1).
 \label{eq:B_giant_joint}
\end{equation}


\subsection{Branch generating functions}

We label retained edges as type~1 and reversed edges as type~2.
Let $g_o^{(\alpha)}(s_o)$ be the probability that following an edge
of type $\alpha$ in its current direction reaches a finite forward
branch containing $s_o$ vertices. The count includes the vertex
reached by that edge and excludes the vertex from which the
traversal started. We define $g_i^{(\alpha)}(s_i)$ analogously for
backward exploration, in which the edge is traversed against its
current direction. Because the original stub matching is uniform, these functions do
not depend on the original degrees of the vertex from which the
edge is followed. They do depend on the edge type, which determines
the original stub through which the next vertex is reached.

These two functions are needed to compute $G_o(s_o\mid\mathbf{k})$ and $G_i(s_i\mid\mathbf{k})$. Consider, for instance the outgoing cluster size distribution of nodes with degree $\mathbf{k}$, $G_o(s_o\mid\mathbf{k})$. This can be written as
\begin{widetext}
\begin{equation}
G_o(s_o\mid\mathbf{k})= \sum_{k'_i=0}^{k_i}\sum_{k'_o=0}^{k_o} \binom{k_i}{k'_i} p^{k'_i} q^{k_i-k'_i}
\binom{k_o}{k'_o} q^{k'_o} p^{k_o-k'_o} \sum_{\{s_n,s_m\}} \prod_{n=1}^{k'_i}\prod_{m=1}^{k'_o} g_o^{(2)}(s_n)g_o^{(1)}(s_m)
\delta_{s_o,1+\sum_{n=1}^{k'_i}s_n+\sum_{m=1}^{k'_o}s_m} .
\label{eq:Go}
\end{equation}
\end{widetext}
From Eq.~\eqref{eq:Go}, we obtain its generating function as
\begin{equation}
\widehat{G}_o(z_o\mid\mathbf{k})=z_o\left(q+p\widehat{g}^{(2)}_o(z_o)\right)^{k_i}\left(p+q\widehat{g}^{(1)}_o(z_o)\right)^{k_o}
\label{eq:GoGenerating}
\end{equation}
and after averaging over $P_0(k_i,k_o)$, we obtain
\begin{equation}
\widehat{G}_o(z_o)=z_o\widehat{P}_0\left(x_f(z_o),y_f(z_o)\right),
\end{equation}
where we have defined the shorthand notation
\begin{equation}
 x_f(z_o)\equiv q+p\widehat g_o^{(2)}(z_o) \; \mbox{and} \; y_f(z_o)\equiv p+q\widehat g_o^{(1)}(z_o)
 \label{eq:B_xfyf}
\end{equation}

Following the same argument used to derive Eqs.~\eqref{eq:Go} and~\eqref{eq:GoGenerating}, we can obtain a closed system of equations for $\widehat g_o^{(1)}(z_o)$ and $\widehat g_o^{(2)}(z_o)$. Arrival through a retained edge reaches an original incoming stub, so the original degrees of the vertex reached are distributed
according to $k_iP_0(\mathbf{k})/\mu$. The arrival edge has already
been used, leaving $k_i-1$ original incoming edges and $k_o$
original outgoing edges. Thus,
\begin{widetext}
\begin{equation}
 \widehat g_o^{(1)}(z_o)
 =
 \frac{z_o}{\mu}
 \sum_{\substack{\mathbf{k}\\ k_i\geq1}}
 k_iP_0(\mathbf{k})
 \left(q+p\widehat{g}^{(2)}_o(z_o)\right)^{k_i-1}\left(p+q\widehat{g}^{(1)}_o(z_o)\right)^{k_o}
 \nonumber\\
 =
 \frac{z_o}{\mu}
 \Pgf_{0,x}\bigl(x_f(z_o),y_f(z_o)\bigr).
 \label{eq:B_forward_retained}
\end{equation}
\end{widetext}
Here, $\Pgf_{0,x}$ denotes the derivative of $\Pgf_0$ with respect
to its first argument whereas the factor $z_o$ takes into account the vertex reached.

Arrival through a reversed edge instead reaches an original
outgoing stub. Its degree distribution is
$k_oP_0(\mathbf{k})/\mu$, and the remaining original degrees are
$k_i$ and $k_o-1$. Therefore,
\begin{widetext}
\begin{equation}
 \widehat g_o^{(2)}(z_o)
 =
 \frac{z_o}{\mu}
 \sum_{\substack{\mathbf{k}\\ k_o\geq1}}
 k_oP_0(\mathbf{k})
 \left(q+p\widehat{g}^{(2)}_o(z_o)\right)^{k_i}\left(p+q\widehat{g}^{(1)}_o(z_o)\right)^{k_o-1}
 \nonumber\\
 =
 \frac{z_o}{\mu}
 \Pgf_{0,y}\bigl(x_f(z_o),y_f(z_o)\bigr).
 \label{eq:B_forward_reversed}
\end{equation}
\end{widetext}
where $\Pgf_{0,y}$ is the derivative with respect to the second
argument.

For backward exploration, an original incoming edge can be
traversed against its current direction if it is retained, whereas
an original outgoing edge can be traversed in this way if it is
reversed. Repeating the same analysis above in this case leads to
\begin{align}
 \widehat g_i^{(1)}(z_i)
 &=
 \frac{z_i}{\mu}
 \Pgf_{0,y}\bigl(x_b(z_i),y_b(z_i)\bigr),
 \nonumber\\
 \widehat g_i^{(2)}(z_i)
 &=
 \frac{z_i}{\mu}
 \Pgf_{0,x}\bigl(x_b(z_i),y_b(z_i)\bigr),
 \label{eq:B_backward_branches}
\end{align}
where we define
\begin{equation}
 x_b(z_i)\equiv p+q\widehat g_i^{(1)}(z_i)
\; \mbox{and}\;
 y_b(z_i)\equiv q+p\widehat g_i^{(2)}(z_i).
 \label{eq:B_backward_factors}
\end{equation}
Finally,
\begin{equation}
\widehat{G}_i(z_i\mid\mathbf{k})=z_i\left(p+q\widehat{g}^{(1)}_i(z_i)\right)^{k_i}\left(q+p\widehat{g}^{(2)}_i(z_i)\right)^{k_o}
\label{eq:GiGenerating}
\end{equation}
and after averaging over $P_0(k_i,k_o)$, we obtain
\begin{equation}
\widehat{G}_i(z_i)=z_i\widehat{P}_0\left(x_b(z_i),y_b(z_i)\right),
\end{equation}

\subsection{Extinction probabilities and the giant in- and out-components}

We now evaluate the branch generating functions at unity and define
\begin{align}
 u&=\widehat g_o^{(1)}(1),
 &
 v&=\widehat g_o^{(2)}(1),
 \nonumber\\
 \bar u&=\widehat g_i^{(1)}(1),
 &
 \bar v&=\widehat g_i^{(2)}(1).
 \label{eq:B_extinction_definitions}
\end{align}
These quantities are the probabilities that the corresponding
forward or backward branches are finite.

Writing $x_f=x_f(1)$ and $y_f=y_f(1)$, the forward equations become
\begin{align}
 x_f&=q+pv,
 &
 y_f&=p+qu,
 \nonumber\\
 u&=\frac{\Pgf_{0,x}(x_f,y_f)}{\mu},
 &
 v&=\frac{\Pgf_{0,y}(x_f,y_f)}{\mu}.
 \label{eq:forward}
\end{align}
Similarly, with $x_b=x_b(1)$ and $y_b=y_b(1)$, the backward
equations are
\begin{align}
 x_b&=p+q\bar u,
 &
 y_b&=q+p\bar v,
 \nonumber\\
 \bar u&=\frac{\Pgf_{0,y}(x_b,y_b)}{\mu},
 &
 \bar v&=\frac{\Pgf_{0,x}(x_b,y_b)}{\mu}.
 \label{eq:backward}
\end{align}
The physical solution is the componentwise smallest fixed point in
$[0,1]^2$ for each pair of equations. It can be obtained by
iteration starting from zero.

Substituting the root generating functions into
Eq.~\eqref{eq:B_giant_marginals} gives
\begin{equation}
 S_{\GIN}=1-\Pgf_0(x_f,y_f),
 \qquad
 S_{\GOUT}=1-\Pgf_0(x_b,y_b).
 \label{eq:ginout}
\end{equation}

Following Ref.~\citep{Boguna2005}, the branching threshold can also
be recovered by linearizing the fixed-point equations around the
trivial solution. The Jacobian of the forward equations at
$(u,v)=(1,1)$ is
\begin{equation}
 M_f
 =
 \frac{1}{\mu}
 \begin{pmatrix}
 qF_{io} & pF_i\\
 qF_o   & pF_{io}
 \end{pmatrix}
 =
 M(p),
 \label{eq:B_forward_jacobian}
\end{equation}
which coincides with Eq.~\eqref{eq:matrix}. The backward equations
give
\begin{equation}
 M_b
 =
 \frac{1}{\mu}
 \begin{pmatrix}
 qF_{io} & pF_o\\
 qF_i   & pF_{io}
 \end{pmatrix}.
 \label{eq:B_backward_jacobian}
\end{equation}
The two matrices have the same trace and determinant and therefore
the same eigenvalues. This recovers the common forward and backward
branching threshold derived in the main text.

\subsection{Joint generating function and the strongly connected component}

To compute the $\GSCC$, we need the joint probability that both
explorations are finite. Since every edge is purely directed, an
incident edge can initiate only one of the two explorations for
any realized orientation. The orientation of that edge nevertheless
enters both component-size calculations and must be averaged
consistently.

To make this explicit, consider a root with original degrees
$(k_i,k_o)$ and use the variables introduced in
Eq.~\eqref{eq:bin}: $n$ is the number of original incoming edges
that are retained, and $m$ is the number of original outgoing
edges that are reversed. They are independent and satisfy
\begin{equation}
 n\sim\operatorname{Binomial}(k_i,q),
 \qquad
 m\sim\operatorname{Binomial}(k_o,p).
 \label{eq:B_orientation_counts}
\end{equation}
The root then has $k_o-m$ retained outgoing edges and $k_i-n$
reversed outgoing edges, together with $n$ retained incoming edges
and $m$ reversed incoming edges.

Conditional on these counts, the forward and backward explorations
use disjoint incident edges. Independence of their branches in the
locally tree-like limit therefore gives
\begin{align}
 &\widehat G(z_i,z_o\mid\mathbf{k},n,m)
 \nonumber\\
 &\quad=
 z_iz_o\,
 \left[\widehat g_o^{(1)}(z_o)\right]^{k_o-m}
 \left[\widehat g_o^{(2)}(z_o)\right]^{k_i-n}
 \nonumber\\
 &\qquad\quad\times
 \left[\widehat g_i^{(1)}(z_i)\right]^n
 \left[\widehat g_i^{(2)}(z_i)\right]^m.
 \label{eq:B_joint_conditioned}
\end{align}
The factor $z_iz_o$ counts the root once in each component.

Averaging over $n$ assigns each original incoming edge the joint
generating factor
\begin{equation}
 q\widehat g_i^{(1)}(z_i)
 +
 p\widehat g_o^{(2)}(z_o).
 \label{eq:B_joint_incoming_factor}
\end{equation}
If that edge is retained, it contributes to the backward exploration;
if it is reversed, it contributes to the forward exploration.
Similarly, averaging over $m$ assigns each original outgoing edge
the factor
\begin{equation}
 q\widehat g_o^{(1)}(z_o)
 +
 p\widehat g_i^{(2)}(z_i).
 \label{eq:B_joint_outgoing_factor}
\end{equation}

Taking the product of these factors over the original incident
edges and then averaging over $P_0(\mathbf{k})$ yields
\begin{align}
 \widehat G(z_i,z_o)
 ={}&
 z_iz_o\,\Pgf_0\Bigl(
 q\widehat g_i^{(1)}(z_i)
 +p\widehat g_o^{(2)}(z_o),
 \nonumber\\
 &\hspace{3.7em}
 q\widehat g_o^{(1)}(z_o)
 +p\widehat g_i^{(2)}(z_i)
 \Bigr).
 \label{eq:B_joint_generator}
\end{align}

Thus, the factorization into forward and backward contributions
holds when the retained and reversed incident-edge counts are
specified. It generally does not survive the average over those
counts at fixed original degrees. The reason is that the same
orientation variable determines whether an original edge contributes
to the forward or to the backward exploration. This dependence is
fully accounted for by the sums in
Eqs.~\eqref{eq:B_joint_incoming_factor}
and~\eqref{eq:B_joint_outgoing_factor}. 

At $z_o=z_i=1$, Eq.~\eqref{eq:B_joint_generator} becomes
\begin{equation}
 \widehat G(1,1)
 =
 \Pgf_0(q\bar u+pv,\,qu+p\bar v).
 \label{eq:B_joint_extinction}
\end{equation}
Finally, substituting this expression and the two marginal
generating functions into Eq.~\eqref{eq:B_giant_joint} gives
\begin{align}
 S_{\GSCC}
 ={}&
 1-\Pgf_0(x_f,y_f)-\Pgf_0(x_b,y_b)
 \nonumber\\
 &+\Pgf_0(q\bar u+pv,\,qu+p\bar v).
 \label{eq:gscc}
\end{align}

As a consistency check, at $p=0$ only retained edges are present.
The relevant extinction probabilities then satisfy
\begin{equation}
 u=\frac{\Pgf_{0,x}(1,u)}{\mu},
 \qquad
 \bar u=\frac{\Pgf_{0,y}(\bar u,1)}{\mu},
 \label{eq:B_unperturbed_extinction}
\end{equation}
and Eq.~\eqref{eq:gscc} reduces to
\begin{align}
 S_{\GSCC}(0)
 ={}&
 1-\Pgf_0(1,u)-\Pgf_0(\bar u,1)
 \nonumber\\
 &+\Pgf_0(\bar u,u),
 \label{eq:B_unperturbed_gscc}
\end{align}
recovering the result for the original directed configuration model.


\end{document}